\documentclass[12pt]{article}
\usepackage{amsmath, amssymb}
\usepackage{amsfonts}
\usepackage{theorem}
\usepackage{makeidx}

{\theorembodyfont{\itshape}}

\newtheorem{theorem}{Theorem} [section]

\newtheorem{lemma}{Lemma}[section]

\numberwithin{equation}{section}

\newcommand{\R}{\mathbb{R}}
\newcommand{\Z}{\mathbb{Z}}

\newcommand{\myproof}{\noindent {\bf Proof}: \quad}

\newcommand{\myendproof}{\hspace*{\fill}{{\bf \small Q.E.D.}} \vspace{10pt}}

\newcommand{\donothing}[1]{{}}

\newcommand{\xb}{\mathbf{x}}

\newcommand{\pb}{\mathbf{p}}

\newcommand{\rb}{\mathbf{r}}
\renewcommand{\sb}{\mathbf{s}}

\newcommand{\U}{{\mathbf U}}
\newcommand{\V}{{\mathbf V}}
\newcommand{\E}{{\cal E}}

\newcommand{\A}{{\mathcal A}}
\newcommand{\calS}{{\mathcal S}}

\newcommand{\trueself}{{\Omega}}

\newcommand{\no}{\nonumber}

\donothing{
\newcommand{\xb}{{x}}

\newcommand{\pb}{{p}}

}

\newcommand{\good}{{\cal G}}
\newcommand{\bad}{{\cal B}}

\newcommand{\goodone}{{ G}}
\newcommand{\badone}{{B}}

\newcommand{\logl}{|\log  \lambda |}
\newcommand{\loggl}{|\log \log \lambda |}
\newcommand{\logggl}{|\log \log \log  \lambda |}

\newcommand{\selfnonekappa}
{\bomega(\rb_{n+1}) |\log (\lambda+ \bomega(\pb_{n+1}))|^\kappa}

\newcommand{\selfnkappa}
{\bomega(\rb_{n}) |\log (\lambda+ \bomega(\pb_{n}))|^\kappa}

\newcommand{\selfnone}
{\bomega(\rb_{n}) |\log (\lambda+ \bomega(\pb_{n}))|}

\newcommand{\rselfnkappatilde}
{\omega(r_{n}) |\log (\lambda+ \bomega(\pb_{n}))|^{1-\kappa/2}}

\newcommand{\rselfnone}
{\omega(r_{n}) |\log (\lambda+ \bomega(\pb_{n}))|}

\newcommand{\Th}{\Theta_\kappa}

\newcommand{\Thnone}
{\Theta_\kappa(\pb_{n+1})}

\newcommand{\diagtkn}{{\cal K}^{\kappa, \good^\tau}_n}
\newcommand{\offonetkn}{{\Phi}^{\kappa, \good^\tau}_n}
\newcommand{\offtwotkn}{{\Psi}^{\kappa, \good^\tau}_n}
\newcommand{\diagtwotkn}{{\cal K}^{\kappa, \good^{2\tau}}_n}
\newcommand{\offonetwotkn}{{\Phi}^{\kappa, \good^{2\tau}}_n}
\newcommand{\offtwotwotkn}{{\Psi}^{\kappa, \good^{2\tau}}_n}
\newcommand{\bomega}{{\boldsymbol {\omega}}}

\renewcommand{\l}{\langle \!\langle}
\renewcommand{\r}{\rangle \!\rangle}
\newcommand{\la}{\langle}
\newcommand{\ra}{\rangle}

\begin{document}

\title{ $(\log t)^{2/3}$ law of the two dimensional asymmetric
simple exclusion process}

\author{
Horng-Tzer Yau\footnote{Work partially supported by NSF grant
DMS-0072098,  yau@cims.nyu.edu} \\ \vspace*{-0.3cm} \\ Courant
Institute,  New York University} 


\maketitle


\begin{abstract}
We prove that the diffusion coefficient for
the two dimensional asymmetric simple exclusion process
diverges as $(\log t)^{2/3}$ to the leading order. The method
applies to nearest and non-nearest neighbor
asymmetric simple exclusion processes.
\end{abstract}


\section{Introduction}

Asymmetric simple exclusion process is a
Markov process on $\{0,1\}^{\mathbb Z^d}$ with asymmetric jump rates.
There is at most one particle allowed per site and thus the name
exclusion.
The particle at a site $x$ waits for an exponential time and then
jump to $y$ with rate $p(x-y)$ provided that the site is not occupied.
Otherwise the jump is suppressed  and the process starts again.
The jump rate is assumed to be asymmetric so that in general there is
net drift  of the system. The simplicity of the model
has made it the default stochastic
model for  transport phenomena.
Furthermore, it is also a basic
component for models \cite{EMY} \cite{QY}
with incompressible Navier-Stokes equations
as the hydrodynamical equation.

The hydrodynamical limit
of the asymmetric simple exclusion process was proved by Rezakhanlou
\cite{R}
to be a viscousless  Burgers equation in the Euler scaling limit.
If the system is in equilibrium, the Burgers equation is trivial
and the system moves with a uniform velicity.  This unifrom
velocity can be removed and the viscosity of the system, or the
diffusion coefficient,
can be defined via the standard mean square displacement.
Although the diffusion coefficient is expected to be finite
for dimension $d >2$, a rigorous proof was obtained only
a few years ago  \cite{LY} by estimating the corresponding
resolvent equation.
Based on the mode coupling theory,
Beijeren, Kutner and  Spohn \cite{Bei Ku Spo}
conjectured that $D(t)\sim
(\log t)^{2/3}$ in dimension $d=2$ and
$D(t)\sim t^{1/3}$ in $d=1$. The conjecture at $d=1$ was also
made by Kardar-Parisi-Zhang via the KPZ equation.

This problem has received much attention recently in the context
of integrable systems. The main quantity analyzed
is fluctuation of the current
across the origin in
$d=1$ with the jump  restricted to the nearest right site, the
totally asymmetric simple exclusion process (TASEP).
Consider
the special configuration that all sites to the left of the origin
were occupied while all sites to the right
of the origin were empty.
Johansson \cite{Joh} observed that
the current across the origin
with this special initial data can be mapped
into a last passage percolation problem.
By analyzing  resulting percolation
problem asymptotically in the limit
$N \to \infty$, Johansson proved that the variance of
the current is of order  $t^{2/3}$.
In the case of discrete time,
Baik and Rains \cite{BR} analyze an extended version of the last passage
percolation problem and  obtain fluctuations of order $t^\alpha$,
where $\alpha =1/3$ or $\alpha =1/2$ depending on the parameters of
the model.
Both the approaches of  \cite{Joh} and
\cite{BR} are related to the earlier results of
Baik-Deift-Johansson \cite{Ba dei jo} on the distribution
of the length of the longest increasing subsequence in random
permutations.

In \cite{PS} (see also \cite{Prae Spo2}),  Pr\"{a}hofer and Spohn
succeeded  in mapping
the current of the TASEP into a last passage percolation problem
for a general class of initial
data, including the equilibrium case considered in this article.
For the discrete time case, the extended problem is
closely related to the work
\cite{BR}, but the boundary conditions
are  different. For continuous time, besides the
boundary condition issue, one has to  extend
the result of \cite{BR}
from the geometric to
the exponential distribution.

To relate these results to our problem,
we consider the asymmetric simple exclusion process in equilibrium
with a Bernoulli product measure of
density $\rho$ as the invariant measure.
Define the time dependent correlation function in equilibrium by
$$
S(x, t)=
\langle  \eta_x(t);
\eta_0(0) \rangle
$$
 We shall choose $\rho=1/2$ so
that
there is not net global drift, $\sum_x x S(x, t) = 0$. Otherwise a
subtraction of the drift should be performed. The diffusion coefficient we
consider is (up to a constant) the second moment of $S(x, t)$:
$$
\sum_x x^2 S(x, t) \sim D(t) t
$$
for large $t$. On the other hand
the variance of the current across the origin
is proportional to
\begin{equation}\label{0.1}
\sum_x |x| S(x, t)
\end{equation}
Therefore, Johansson's result on the variance of
the current  can be interpreted
as the spreading of  $S(x,t)$ being of order $t^{2/3}$.
The result of Johansson is for special initial data and does not directly
apply to the equilibrium case.
If we combine the work of \cite{PS}
and \cite{BR}, neglect various issues discussed above, and extrapolate
to the second moment, we obtain a growth of the second moment as
$t^{4/3}$, consistent with the conjectured $D(t) \sim t^{1/3}$.

We remark that the results based on integrable
systems are not just for the variance of the current
across the origin, but also for its full limiting distribution.
The main restrictions
appear to be the rigid requirements of the fine details of the dynamics
and the initial data. Furthermore, it is not clear whether the analysis
on the current across the origin can be extended to the diffusivity.
In particular, the divergence of
$D(t)$  as $t \to \infty$ in $d=1$ has not been proved
via this approach even for the
TASEP.

Recent work of \cite{SY} has taken a completely different approach.
It is based on the analysis of the Green
function of the dynamics. One first uses
the duality to map the resolvent equation into a system
of infinitely-coupled equations. The hard core condition
was then removed by using variational principles.
Once the hard core condition was removed,
the Fourier transform becomes a very useful tool and the Green function was
estimated  to degree three. This yields a lower
bound to the full Green function via a monotonicity inequality.
Thus one obtains the lower bounds
$D(t) \ge t^{1/4}$ in $d=1$
and $D(t) \ge (\log t)^{1/2}$ in $d=2$ \cite{SY}.
In this article, we shall estimate the Green function to degrees
high enough to determine  the leading order behavior
$D(t) \sim (\log t)^{2/3}$ in $d=2$.

\bigskip

\subsection{Definitions of the Models}

\bigskip

Denote the configuration by $\eta = (\eta_x)_{x \in \Z^d}$
where $\eta_x=1$ if the site $x$ is occupied
and $\eta_x=0$ otherwise. Denote
$\eta^{x,y}$  the configuration obtained from $\eta$
by exchanging the occupation variables at $x$ and $y$~:
$$
(\eta^{x,y})_z\; =\;
\begin{cases}
\eta_z & \hbox{if $z\neq x$, $y$,}\\
\eta_x & \hbox{if $z=y$ and}\\
\eta_y & \hbox{if $z=x$.}
\end{cases}
$$
Then the  generator  of
the  asymmetric simple exclusion process is given by
\begin{equation}
({\mathcal L}f) (\eta)\; =\; \sum_{j=1}^d \sum_{x\in\Z^d} p(x,y)\eta_x
[1-\eta_{y}] [f(\eta^{x,y}) -f(\eta)]\; .
\label{1g0}
\end{equation}
where $\{e_k, \, 1\le k\le d\}$ stands for the canonical
basis of $\Z^d$.
For each $\rho$ in $[0,1]$, denote by
$\nu_\rho$ the Bernoulli product measure on $\{0,1\}^{\mathbb Z^d}$
with density $\rho$ and by $<\cdot, \cdot>_\rho$ the inner product in
$L^2(\nu_\rho)$. The probability measures $\nu_\rho$ are invariant
for the asymmetric simple exclusion process.

For two cylinder functions $f$, $g$ and
a density $\rho$, denote by $\la f;g\ra_\rho$ the covariance of $f$ and
$g$ with respect to $\nu_\rho$~:
$$
\la f;g\ra_\rho \; =\; \la fg\ra_\rho - \la f\ra_\rho \la g\ra_\rho.
$$
Let $P_\rho$ denote the law of the
asymmetric simple exclusion process starting from the equilibrium
measure $\nu_\rho$.
Expectations with respect to  $P_\rho$ is
denoted by $E_\rho$.
Let
$$
S_\rho(x, t)= E_\rho [ \{ \eta_x(t) -\eta_x(0)\} \eta_0(0)]
=\langle \eta_x(t) ;  \eta_0(0) \rangle_\rho
$$
denote  for the time dependent correlation functions in
equilibrium with density  $\rho$. The compressibility
$$
\chi=\chi (\rho) = \sum_x \la \eta_x ; \eta_0 \ra_\rho
= \sum_x S_\rho(x, t)
$$
is time independent and $\chi (\rho) = \rho (1-\rho)$ in our setting.

The bulk diffusion coefficient is the variance of the position w.r.t. the
probability measure $S_\rho(x, t)\chi^{-1}$ in $\Z^d$ divided by $t$,
i.e.,
\begin{equation}
D_{i,j}(\rho,t)  \; =\; \frac{1}{t}
\bigg\{ \sum_{x\in\Z^d}
x_i x_j S_\rho(x, t)\chi^{-1}
-(v_i t)(v_j t)\bigg\},
\label{0.4}
\end{equation}
where  $v$ in $\R^d$ is the velocity defined by
\begin{equation}
vt\; =\;
\sum_{x\in\Z^d}  x S_\rho(x, t)\chi^{-1}\; .
\label{0.9}
\end{equation}

For simplicity, we shall restrict ourselves
to the case that the jump is symmetric in the $y$ axis but totally asymmetric
in the $x-$axis, i.e., only jump to the right is allowed in the
$x-$axis.  Our results
hold for other jump rates as well.
The  generator of this process is given by
\begin{equation}
({\mathcal L}f) (\eta)\; =\;  \sum_{x\in\Z^d} \Big [ \eta_x
(1-\eta_{x+ e_1 }) (f(\eta^{x,x+ e_1 }) -f(\eta))   +
  \frac 1 2 \big ( f(\eta^{x,x +  e_2 }) -f(\eta)\big )  \Big ]
\label{1g}
\end{equation}
where we have combined the symmetric jump in the $y$-axis into the last term.
We emphasize that the result and method in this paper apply
to all asymmetric simple exclusion processes; the special choice
is made to simplify the notation.
The velocity of the totally asymmetric simple exclusion process
is  explicitly computed as $ v\; =\; 2 (1-2\rho) e_1$.
We further assume that the density is $1/2$ so that the velocity is zero
for simplicity.

Denote the instantaneous currents (i.e., the difference between
the rate at which a particle jumps from $x$ to $x+e_i$ and the
rate at which a particle jumps from $x+e_i$ to $x$) by $\tilde w_{x,x+e_i}$:
\begin{equation}
\tilde w_{x,x+ e_1 }\; =\;  \eta_x [1-\eta_{x+ e_1 }],
\qquad \tilde w_{x, x+e_2} = \frac { \eta_{x+ e_2 } - \eta_x} 2
\label{0.12}
\end{equation}
We have the conservation law
$$
{\mathcal L} \eta_0 + \sum_{i=1}^2  \Big\{ \tilde
w_{-e_i,0}-\tilde w_{0,e_i}\Big\} = 0\; .
$$
Let ${ w}_i(\eta)$ denote
the renormalized current in the $i$-th direction:
\begin{equation*}
 { w}_i(\eta)  = \tilde w_{0,e_i}  - \la \tilde w_{0,e_i}\ra_\rho  -
\frac d{d\theta}  \la \tilde w_{0,e_i}\ra_{\theta}\Big\vert_{\theta=\rho}
(\eta_0 -\rho)
\end{equation*}
Notice  the subtraction of the linear term in this definition.
We have
\begin{equation*}
 { w}_1(\eta)  = (\eta_0 -\rho)(\eta_{e_1 } -\rho)+ \rho [\eta_{e_1} - \eta_0]
, \quad
 { w}_2(\eta) = \frac { \eta_{e_2} - \eta_0 } 2
\end{equation*}

Define the semi-inner product
\begin{equation}
{\l}  g, h {\r}_\rho \; =\;
 \sum_{x\in \mathbb Z^d } <  \tau_x g \, ;  \,  h >_\rho
\; =\;   \sum_{x\in \mathbb Z^d } < \tau_x h \, ;  \,  g >_\rho\;
\cdot \label{1a}
\end{equation}
Since the subscript $\rho$ is fixed to be $1/2$ in this paper, we
shall drop it. All but a finite number of terms in this sum vanish
because $\nu_\rho$ is a product measure and $g$, $h$ are mean
zero. From this inner product, we define the norm:
\begin{equation}
\| f \|^2 = \l  f, f \r.
\label{norm}
\end{equation}

Notice that all degree one functions vanish in this norm and we
shall identify the currents $w$ with their degree two parts.
Therefore, for the rest of this paper, we shall put
\begin{equation}
 { w}_1(\eta)  = (\eta_0 -\rho)(\eta_{e_1 } -\rho)
, \quad
 { w}_2(\eta) = 0
\label{0.8}
\end{equation}

Fix a unit vector  $\xi \in \Z^d$. From some
simple calculation using the Ito's formula \cite{LOY}
we can rewrite the diffusivity as
\begin{equation}
\xi \cdot D \xi  - {\frac 1 2}
 \; = \;
\frac{1}{\chi}
\; \left\|  t^{-1/2}\int_0^t ds\;
(\xi \cdot { w}) (\eta(s) ) \right\|^2.
\label{9.3}
\end{equation}
This is some variant of the  Green-Kubo formula.
Since $w_{2}=0$, $D$ is a matrix with all entries zero except
$$
D_{11}
 \; = \; \frac 12 +
\frac{1}{\chi}
\; \left\|  t^{-1/2}\int_0^t ds\;
 { w}_1 (\eta(s) ) \right\|^2.
$$

Recall that $\int_0^\infty e^{-\lambda t} f(t) dt \sim \lambda^{-\alpha}$
as $\lambda\to 0$ means (in some weak sense) that $f(t)\sim t^{\alpha -1}$.
Throughout the following $\lambda$ will always be a positive real number.
The main result of this article is the following Theorem.


\begin{theorem}\label{th:main}  Suppose that the density $\rho= 1/2$.
Then there exists a constant  $\gamma > 0 $ so that for sufficiently small $\lambda>0$,
$$
\lambda^{-2} \logl^{2/3} e^{ - \gamma \logggl^{2} }
\le \int_0^\infty e^{-\lambda t} \; t D_{11} (t) dt \le \lambda^{-2}
  \logl^{2/3} e^{  \gamma \logggl^{2} }
$$
\end{theorem}

\bigskip

From the definition, we can rewrite the diffusion coefficient as
$$
t  D_{11}(t) = \frac t 2 +  \frac 2 \chi  \int_0^t\int_0^s \;
\l e^{ u {\mathcal L}} w_1,w_1\r \; du ds
$$
Thus
\begin{eqnarray}\label{intid}
\int_0^\infty e^{-\lambda t}\;  t D_{11}(t)  dt
& = & \frac 1 {2 \lambda^2} + \frac 2 \chi \int_0^\infty dt \int_0^t\int_0^s e^{-\lambda t}\;
 \; \l e^{ u {\mathcal L}} w_1,w_1 \r \; du ds  \no \\
& = & \frac 1 {2 \lambda^2} + \frac 2 \chi \int_0^\infty  du  \Big \{
\int_u^\infty d t \;  e^{-\lambda (t-u) }
\Big ( \int_u^t d s \; \Big )\; \Big \}
 \; \l e^{-\lambda u } e^{ u {\mathcal L}} w_1 \; , \; w_1 \r  \no \\
& = & \frac 1 {2 \lambda^2} +\chi^{-1} \lambda^{-2}\l w_1
\; ,\; (\lambda-{\mathcal L})^{-1} w_1\r.
\end{eqnarray}
Therefore, Theorem \ref{th:main} follows from
the following estimate on the resolvent.

\begin{theorem}\label{resest} There exist a constant $\gamma > 0$ such that
for sufficiently small $\lambda>0$,
$$
\logl^{2/3} e^{ - \gamma \logggl^2 }
\le
\l w_1,(\lambda-{\mathcal L})^{-1} w_1\r \le
  \logl^{2/3} e^{ \gamma \logggl^2 }
$$
\end{theorem}

\bigskip
From the following well-known Lemma, the upper bound holds without
the time integration. For a proof, see \cite{LY}.

\begin{lemma}\label{resb}
Suppose $\mu$ is an invariant measure of a process with generator
${\mathcal L}$. Then
\begin{equation}
 E^\mu \Big [ \Big (
        t^{-1/2} \int_0^t w (\eta(s)) \,ds \Big )^2 \Big ]
\le  \;\l w_1, (t^{-1}-{\mathcal L})^{-1} w_1 \r \label{bound}
\end{equation}
\end{lemma}
Since $w_1$ is the only non-vanishing current, we shall drop the
subscript $1$.

\section{Duality and Removal of the Hard Core Condition}

\donothing{ For each positive integer $n$, denote by
$\mathcal  P_n=\mathcal  P_n(\mathbb Z^d)$ the space of all finite subsets
$\Lambda\subset \Z^d$ of cardinality $n$ and by $|\Lambda|$ the
cardinality of a finite subset $\Lambda$ of $\Z^d$.
}

Denote by $\mathcal  C = \mathcal  C (\rho)$ the space of
$\nu_\rho$-mean zero cylinder functions. For a
finite subset $\Lambda$ of $\mathbb Z^d$, denote by
$\xi_\Lambda$ the mean zero cylinder function
defined by
$$
\xi_\Lambda\; =\;\prod_{x\in\Lambda} \xi_x,
\qquad \xi_x =
\frac { \eta_x-\rho} {\sqrt {\rho(1-\rho)}}\; .
$$
Denote by $\mathcal  M_n$
the space of cylinder homogeneous functions of degree $n$, i.e.,
the space generated by all homogeneous monomials of degree $n$~:
$$
\mathcal  M_n\; =\; \Big\{ h\in\mathcal  C\, ;\;
h = \sum_{ |\Lambda| = n}
h_\Lambda \xi_\Lambda\, , \; h_\Lambda\in\R\Big\}\; .
$$
Notice that in this definition all but a finite number of
coefficients $h_\Lambda$ vanish because $h$ is assumed to
be a cylinder function. Denote by
$\mathcal  C_n=\cup_{1\le j\le n} \mathcal  M_j $ the space
of cylinder functions of degree less than or equal to $n$.
All mean zero cylinder functions $h$ can be decomposed as
a finite linear combination of cylinder functions of finite
degree~: $\mathcal  C = \cup_{n\ge 1} \mathcal  M_n$.
Let ${\mathcal L}=\calS+ \A $ where
$\calS$ is the symmetric part and
$\A$ is the asymmetric part.
Fix a function $g$ in $\mathcal  M_n$~:
$g=\sum_{\Lambda,|\Lambda|=n} g_\Lambda \xi_\Lambda$.
A simple computation shows that the symmetric part is given by
$$
( \calS g)(\eta) \; =\; -\frac{1}{2} \sum_{x\in \mathbb Z^d}
\sum_{\stackrel {\scriptstyle \Omega, \, |\Omega|=n-1}
{\scriptstyle \Omega  \cap \{ x, x+ e_1 \} = \phi}} \Big\{ g_{
\Omega \cup \{ x+ e_1 \} } -  g_{ \Omega \cup \{ x\} } \Big\} \Big
[\xi_{\Omega \cup \{ x+ e_1 \}} - \xi_{\Omega \cup \{ x\}} \Big]\;
.
$$
The asymmetric
part $\A$ is decomposed into two pieces $\A= M+J$
so that $M$ maps $\mathcal  M_n$ into itself and $J=J_+ + J_-$
maps $\mathcal  M_n$ into $\mathcal  M_{n-1}\cup \mathcal  M_{n+1}$:
\begin{align*}
& (M g)(\eta) \; =\; \frac {1-2\rho} 2 \sum_{x\in \Z^d}
\sum_{\stackrel {\scriptstyle \Omega, \, |\Omega|=n-1
}{\scriptstyle \Omega  \cap \{ x, x+ e_1 \} = \phi}} \Big\{ g_{
\Omega \cup \{ x+ e_1 \} } - g_{ \Omega \cup \{ x\} } \Big\}
\Big [\xi_{\Omega \cup \{ x+ e_1 \}} + \xi_{\Omega \cup \{ x\}} \Big]\; , \\
& \qquad (J_+ g) (\eta) \; =\; -\sqrt{\rho(1-\rho)}
\sum_{x\in \Z^d}  \sum_{\stackrel {\scriptstyle \Omega, \, |\Omega|=n-1
}{\scriptstyle \Omega  \cap \{ x, x+ e_1 \} = \phi}}
\Big\{ g_{ \Omega \cup \{ x+ e_1  \} } -  g_{ \Omega \cup \{ x\} } \Big\}
\xi_{\Omega \cup \{x, x+ e_1 \}}\; , \\
& \qquad\qquad  (J_- g) (\eta) \; =\; -\sqrt{ \rho(1-\rho)}
\sum_{x\in \Z^d}  \sum_{\stackrel {\scriptstyle \Omega, \, |\Omega|=n-1
}{\scriptstyle \Omega  \cap \{ x, x+ e_1 \} = \phi}}
\Big\{ g_{ \Omega \cup \{ x+ e_1  \} } -  g_{ \Omega \cup \{ x\} } \Big\}
\xi_{\Omega}\; .
\end{align*}

Restricting to the case $\rho=1/2$, we have $M=0$ and thus  $J=\A$.
Furthermore, $A_+^\ast = -A_-$.
We shall now identify  monomials of degree $n$ with symmetric
functions of $n$ variables. Let $\E_1$ denote the set with no
double sites, i.e.,
$$
\E_1 = \{ \xb_n := (x_1, \cdots, x_n): x_i \not = x_j, \text{ for } i \not = j \}
$$
Define
\begin{align}
f(x_1, \cdots, x_n) & =  f_{\{x_1, \cdots, x_n\}},
\qquad & \text { if } \xb_n \in \E_1  \; \no \\
& =  0,  \qquad & \text  { if }  \xb_n \not \in \E_1.
\end{align}
Notice that
$$
E\big \{ \big [ \sum_{|A| = n} f_A \xi_A \big ] ^2 \big \}
 = \frac 1 {n!} \sum_{x_1, \cdots, x_n \in \Z^d} |f(x_1, \cdots, x_n)|^2
$$
From now on, we shall refer to  $f(x_1, \cdots, x_n)$
as homogeneous function of degree $n$ vanishing on $\E_1$.

With this identification, we have
$$
w_1 (0, e_1)=w_1(e_1,0) =1
$$
and zero otherwise. Since we only have one non-vanishing current, we shall drop
the subscript $1$ for the rest of this paper.

If $g$ is a  symmetric homogeneous function of degree $n$, we can check that
\begin{eqnarray}\label{Afdef}
& & A_+ g(x_1, \cdots,  x_{n+1})  \no\\
& = &  - \frac  1 2 \sum_{ i = 1}^ {n+1} \sum_{j \not =  i }
[g(x_1, \cdots, x_{i}+ e_1, \cdots, \widehat { x_{j}}, \cdots
x_{n+1})-
g(x_1, \cdots, x_{i},\cdots,  \widehat { x_{j}}, \cdots, x_{n+1})] \no \\
& & \times \delta (x_{{j}}- x_{i}-e_1 ) \prod_{k \not = j} \big (\,
1-\delta (x_{j}-x_{k}) \, \big )
\end{eqnarray}
where $\delta (0)=1$ and zero otherwise. We can check that
\begin{align}
  \calS g(x_1, \cdots,  x_{n})
& =  \alpha \sum_{ i = 1}^ {n} \sum_{\sigma = \pm } \sum_{ \alpha = 1, 2}
 \prod_{k \not = i} \big (\, 1-\delta (x_{i} + \sigma e_\alpha-x_{k}) \, \big )
\no \\
&   \times [g(x_1, \cdots x_{i}+ \sigma e_\alpha, \cdots, x_{n})-
 g(x_1, \cdots, x_{i},, \cdots, x_{n})]
\end{align}
where $\alpha$ is some constant and $\delta (0)=1$ and zero otherwise.
The constant $\alpha$ is not important in this paper and we shall fix it
so that $\calS$ is the same as the discrete Laplacian with Neumann
boundary condition on $\E_1$.

\bigskip

The hard core condition
makes various  computation very complicated. In particular,
the Fourier transform is difficult to apply. However,
if we are interested
only in the orders of magnitude, this condition  was
removed in \cite{SY}. We now summarize the main result in \cite{SY}.

For a function $F$, we shall use the same symbol $\langle F
\rangle $ to denote the expectation
$$
\frac 1 {n!} \sum_{x_1, \cdots, x_n \in \Z^2} F(x_1, \cdots, x_n)
$$
We now define $A_+F $ using the same formula except we drop the last delta
function, i.e,
\begin{eqnarray}\label{AFdef}
{ A}_+ F(x_1, \cdots,  x_{n+1}) &  = & - \frac 1 2  \sum_{ i = 1}^
{n+1} \sum_{j \not =  i } \Big [ \; F(x_1, \cdots, x_{i}+ e_1 ,
\cdots, \widehat { x_{j}}, \cdots  x_{n+1})  \no \\
&& \; - F(x_1, \cdots, x_{i},\cdots, \widehat {
x_{j}}, \cdots,  x_{n+1})\; \Big ] \,  \delta (x_{{j}}- x_{i}-e_1 )
\end{eqnarray}
Notice that $\langle A_+ F \rangle = 0$. Thus the counting
measure is invariant  and we  define $A_- = - A_+^\ast$, i.e.,
\begin{equation}
\langle A_- G, F\rangle = -\langle  G, A_+ F\rangle
\end{equation}
Finally,
we define
$$
L=\Delta + A, \qquad A=A_+ + A_-,
$$
where the discrete Laplacian is given by
\begin{eqnarray*}
  \Delta F(x_1, \ldots,  x_{n})
 =  \sum_{ i = 1}^ {n} \sum_{\sigma = \pm } \sum_{ \alpha = 1, 2}
    [F(x_1, \ldots x_{i}+ \sigma e_\alpha, \ldots, x_{n})-
 F(x_1, \ldots, x_{i},, \ldots, x_{n})]
\end{eqnarray*}
For the rest of this
paper, we shall only work with $F$ and $L$. So all functions are
defined everywhere and $L$ has no hard core condition.

Denote by $\pi_n$ the projection onto functions with degrees less than or equal to $n$.
Let $L_n$ be the projection  of $L$ onto the image of $\pi_n$, i.e.,
$L= \pi_n L \pi_n$. The key result of \cite{SY} is the
following Lemma.

\bigskip
\begin{lemma}\label{le:mono}
For any $\lambda > 0$ fixed, we have for $k \ge 1$,
\begin{equation}\label{mono3}
C^{-1}  k^{-10} \l w, L_{ 2k+1}^{-1} w\r \le
\l w, {\mathcal L}^{-1} w \r\le  C k^5 \l w, L_{2k}^{-1} w\r
\end{equation}
\end{lemma}
\bigskip

The expression  $\l w, L_{n}^{-1} w\r$ was also calculated in
\cite{SY}. The resolvent equation $(\lambda - L_n) u = w$ can be
written as
\begin{align}\label{truncatedN}
(\lambda-{S}) u_n -A_+ u_{n-1}= & 0,  \no \\
A_+^* u_{k+1} + (\lambda-{S}) u_k -A_+ u_{k-1}= & 0, \quad n-1\ge
 k \ge 3 \\
A_{+}^* u_{3} + (\lambda-S) u_2 = & w.  \no
\end{align}
We can solve  the first  equation  of \eqref{truncatedN} by
$$
 u_n = (\lambda-S)^{-1} A_{+} u_{n-1}.
$$
Substituting this into the equation of degree $n-1$, we have
$$
u_{n-1} = \Big [ (\lambda-S)  + A_{+}^\ast  (\lambda-S) ^{-1}
A_{+} \Big]^{-1} u_{n-2}.
$$
Solving iteratively we arrive at
$$
u_2 =
 \Big [  (\lambda-S) + A_{+}^\ast \Big \{  (\lambda-S)
 + \cdots +A^*_+\Big ( (\lambda-S)
+ A_{+}^\ast  (\lambda-S)^{-1} A_{+} \Big )^{-1} A_{+} \Big
\}^{-1}
 A_{+} \Big ]^{-1} w.
$$
This gives an explicit expression for $\l w, (\lambda-L_3)^{-1} w \r$, for example,
\begin{align}\label{egg}
& \l w, (\lambda-L_3)^{-1} w \r  = \l w, \Big [  \lambda-S  +
A_{+}^\ast (\lambda-S)^{-1} A_{+} \Big ]^{-1} w \r .  \no \\
& \l w, (\lambda-L_4)^{-1} w \r  = \l w, \Big [  \lambda-S +
A_{+}^\ast \Big \{ \lambda-S+ A_{+}^\ast
 (\lambda-S)^{-1}A_{+} \Big \}^{-1} A_{+} \Big ]^{-1} w\r . \no \\
& \l w, (\lambda-L_5)^{-1} w \r  \\& = \l w, \Big [  \lambda-S +
A_{+}^\ast \Big \{ \lambda-S+ A_{+}^\ast
 [\lambda-S+A_{+}^\ast (\lambda-S)^{-1} A_{+} ]^{-1}A_{+} \Big \}^{-1} A_{+} \Big ]^{-1} w\r . \no
\end{align}
This is the expression we shall estimate for the rest of the
paper.

\section{Main Estimate}
\bigskip

We now introduce the following convention: Denote the component of
$p$ by $(r, s)$. Denote $\pb_n = (p_1, \cdots, p_n),  \rb_n = (r_1, \cdots, r_n)$
and $\sb_n= (s_1, \cdots, s_n)$. The Fourier transform of
\[
 [F( x _1+ e_1 , \cdots,  x _n)-F( x _1,\cdots,   x _n)]
 \delta ( x _{n+1}=  x _1+ e_1  )
\]
is given by
\begin{align*}
&  \sum_ x  [F( x _1+ e_1 , \cdots,  x _n)
-F( x _1,\cdots,   x _n)]e^{- i [ x _1  p _1+\cdots+   x _n  p _n
+( x _1+ e_1 )  p _{n+1} ] } \\
&
=  {\left [e^{ i  r_1 }- e^{- i  r_{n+1} } \right ]}
\; \hat F( p _1+ p _{n+1}, \cdots,  p _{n})
\sim  \left [  i ( r_1+  r_{n+1})  \right ] \;
\hat F( p _1+ p _{n+1}, \cdots,  p _{n})
\end{align*}
All functions considered for the rest of this paper
are symmetric periodic functions of period $2 \pi$.

\donothing{
 We now compute the Fourier transform of
\[
 [F( x _1+ e _1, \cdots,  x _n)-F( x _1,\cdots,   x _n)]
 \delta_{n+1}( x _{n+1}-  x _1+ e _1)
\]
By definition, it is
$$
 \sum_ x  [F( x _1+ e _1, \cdots,  x _n)
-F( x _1,\cdots,   x _n)]e^{- i [ x _1  p _1+\cdots+   x _n  p _n
+( x _1+ e _1)  p _{n+1} ] }
$$
$$
=
  {e^{- i  e _1  p _{n+1} }} \sum_ x
[F( x _1+ e _1, \cdots,  x _n) -F( x _1,\cdots,   x _n)] e^{- i [
x _1 ( p _1+ p _{n+1})+\cdots+   x _n  p _n ] }
$$
$$
=  {e^{- i  e _1  p _{n+1} }} \sum_ x F( x _1,\cdots,   x _n)
\left [e^{- i [( x _1- e _1) ( p _1+ p _3)+  x _2  p _2] }- e^{- i
[ x _1 ( p _1+ p _{n+1})+\cdots+   x _n  p _n ] } \right ]
$$
$$
=  {e^{- i  e _1  p _{n+1} }} \left [e^{ i  e _1 ( p _1+ p _{n+1})
}- 1 \right ] \; \hat F( p _1+ p _{n+1}, \cdots,  p _{n+1})
$$
$$
=  {\left [e^{ i  e _1  p _1}- e^{- i  e _1  p _{n+1} } \right ]}
\; \hat F( p _1+ p _{n+1}, \cdots,  p _{n+1})
$$
$$
\sim  \left [  i  e _1 ( p _1+  p _{n+1})  \right ] \; \hat F( p
_1+ p _{n+1}, \cdots,  p _{n+1})
$$
}

Since $F$ is symmetric,  we have
\begin{equation}\label{def:AF}
\widehat { A_+ F} ( \pb _{n+1}) =    - \sum_{j <  m}^{n+1} (e^{i
r_j}- e^{-i r_m}) \hat{ F}( p_1,  \cdots, p_j+p_{m}, \cdots,
\widehat{p_m}, \cdots, p_{n+1})
\end{equation}
We can also compute the discrete Laplacian acting on $F$
$$
\widehat {-\Delta F} ( \pb _{n}) =  - \sum_{j=1}^n \sum_{k=1, 2}
\left [e^{ i  e _k  p _j}-2+  e^{- i  e _k  p _{j} } \right ] \;
\hat F( p _1, \cdots,  p _{n}) = {\bomega}(\pb_n) \hat F( \pb _{n})
$$
where $ \bomega(\pb_n) = \sum_{j=1}^n \omega(p_j)$ and
\begin{equation}
\omega(p_j) = -\left [e^{ i  r_j}-2+  e^{- i  r_{j} } \right ]
-\left [e^{ i  s_j}-2+  e^{- i  s_{j} } \right ]
\end{equation}
We shall abuse the notation a bit by denoting also
$$
\omega(r_j)= -\left [e^{ i  r_j}-2+  e^{- i  r_{j} } \right ]
$$
Notice that $\sqrt {\omega(x)} = 0$ iff $x\equiv 0 $ mod $ \pi$. When
$x\sim 0 $ mod $\pi$, we have
\begin{equation}\label{omegasize}
\sqrt {\omega(x)} \sim |\sin{x}|
\end{equation}

By definition, we have
\begin{align}
\l  F,   G  \r
& =  \sum_z \frac 1 {n!} \sum_{x_1, \cdots, x_n}
\bar F(x_1, \cdots, x_n) G(x_1+z, \cdots, x_n+z) \no \\
& = \sum_z \frac 1 {n!} \int dp_1 \cdots d p_n
\overline{ \hat F (p_1, \cdots, p_n)}  \hat G(p_1,
\cdots, p_n) e^{i (p_1+\cdots+ p_n) z} \no \\
& =  \frac 1 {n!} \int dp_1 \cdots d p_n
\delta\big (p_1+\cdots+ p_n\big )\overline{ \hat F (p_1, \cdots, p_n)} \hat G (p_1, \cdots, p_n)
\end{align}

In other words, when consider the inner product $\l \cdot, \cdot \r$, we
can consider the class of $\hat F(p_1, \cdots, p_n)$ defined
only on the subspace $\sum_j p_j\equiv 0 $ mod $2 \pi$. We shall
simply use the notation $\sum_j p_j = 0$ to denote the last condition.

From now on, we work only on the moment space and all functions
are defined in terms of the momentum variables.
Let $d \mu_{n+1} (\pb_{n+1})$ denote the measure
\begin{equation}\label{mudef}
d \mu_{n+1} (\pb_{n+1}) = \frac 1 { (n+1)!} \,
\delta\, \Big ({\sum_{j=1}^{n+1} p_j}\Big )
  \prod_{j=1}^{n+1} d p_j
\end{equation}

\bigskip

\subsection{Statement of the Main Estimate}
\bigskip
Let $\tau$ be a positive constant and define
\begin{equation}
\good^\tau (\pb_n) =\{\, \bomega(\pb_n)\le |\log \lambda|^{-2\tau} \; \}
\end{equation}
Denote the
complement of $\good^\tau$ by $\bad^\tau$.
Define for $\kappa \ge 0$ the two operators
\begin{align}
\U^n_{\kappa, \tau} (\pb_n) & = \selfnkappa, \; & \pb_n \in \good^\tau ;  \no \\
& = \selfnone, \;
& \pb_n \in \bad^\tau  \label{def:U} \\
\V^n_{\kappa, \tau} (\pb_n)& = \selfnkappa, \; &
\pb_n \in \good^\tau;  \no \\
 &
 =-|\log \log \lambda|^{2}  \bomega(\rb_n) , \;
 &  \pb_n \in \bad^\tau .
\label{def:V}
\end{align}
The main estimates of this paper are the following Theorem.

\begin{theorem} \label{mainest}
Let $\kappa$ and $\tau$ be nonnegative numbers satisfying
\begin{equation}\label{mbound}
 0\le \kappa \le 1  < \tau
\end{equation}
Let $n$  be any positive integers
such that
\begin{equation}\label{nbound}
n^{10} \le \loggl^{1/2}
\end{equation}
Suppose that for some $\gamma \le \loggl^{-3}$
\begin{equation}\label{lowassumption}
 \trueself_{n+1} \ge \gamma \V^{n+1}_{\kappa, 2 \tau}
\end{equation}
as an operator. Let
\begin{equation}
\tilde \kappa= {1-\kappa/2}\; .
\end{equation}
Then
\begin{equation}\label{upest}
  A_+^* (\lambda- S_{n+1}+ \trueself_{n+1})^{-1} A_+ \le
\gamma^{-1} |\log \log \lambda|^2 \U^n_{\tilde  \kappa,  \tau}
\end{equation}
as an operator.

On the other hand, if
\begin{equation}\label{upassumption}
 \trueself_{n+1} \le \gamma^{-1} \U^{n+1}_{\kappa,  \tau}
\end{equation}
Then we have
\begin{equation}\label{downest}
  A_+^* (\lambda- S_{n+1}+ \trueself_{n+1})^{-1} A_+ \ge
C \gamma \V^{n}_{\tilde \kappa, 2 \tau}
\end{equation}
as an operator.
\end{theorem}

\bigskip

\section{Upper Bound}

We first recall that for any two positive operators $A, B$, we have
$$
0< A \le B  \quad \text { iff } \quad 0 < B^{-1} \le A^{-1}\; .
$$
Furthermore, the map $ B \to C^\ast B C $ is monotonic.
For $\gamma \le |\log \log \lambda|^{-3} $, we have
$$
\bomega(\pb_{n+1})+ \gamma \V^{n+1}_{\kappa,
2\tau}(\pb_{n+1})
\ge 0
$$
Thus  we can
substitute the $\Omega$ in Theorem \ref{mainest} by either
$\V$ or $\U$ in the proof.

By definition,
\begin{equation}
 \l F, \;  A^* (\lambda- S_{n+1}+ \gamma \V^{n+1}_{\kappa, 2\tau})^{-1} A  F \r
= \int\;  d \mu_{n+1} (\pb_{n+1}) \;  \frac {| { A_+ F} ( p _1,
\cdots,  p _{n+1})|^2} {\lambda+ \bomega(\pb_{n+1})+ \gamma
\V^{n+1}_{\kappa, 2\tau}(\pb_{n+1}) }  \no
\label{up1}
\end{equation}
Let $\V^{\pm, n+1}_{\kappa, 2\tau}$ denote the
positive and negative parts of $\V^{n+1}_{\kappa, 2\tau}$.  Then
$$
\lambda+ \bomega(\pb_{n+1})+ \gamma \V^{n+1}_{\kappa,
2\tau}(\pb_{n+1}) \ge (1-\gamma)\bomega(\pb_{n+1})+ \gamma \V^{-,
n+1}_{\kappa, 2\tau}(\pb_{n+1}) +\gamma \big [ \lambda+ \bomega
(\pb_{n+1})+ \V^{n+1}_{\kappa, 2\tau}(\pb_{n+1})]
$$
Since $\gamma \le |\log \log \lambda|^{-3} $, we have
$$
(1-\gamma)\bomega(\pb_{n+1})+ \gamma \V^{-, n+1}_{\kappa,
2\tau}(\pb_{n+1}) \ge 0 \; .
$$
Thus we have
\begin{align}
& \l F, \;  A^* (\lambda- S_{n+1}+ \gamma \V^{n+1}_{\kappa, 2\tau})^{-1} A  F \r  \no \\
& \le \gamma^{-1}  \int\;  d \mu_{n+1} (\pb_{n+1}) \;  \frac {|
{ A_+ F} ( p _1, \cdots,  p _{n+1})|^2} {\lambda+ \bomega(\pb_{n+1})+
\V^{+, n+1}_{\kappa, 2\tau}(\pb_{n+1}) } \; .
\label{up12}
\end{align}

We now divide the integration into the good region
$\good^{2\tau}(\pb_{n+1})$ and the bad region
$\bad^{2\tau}(\pb_{n+1}) $.
In the good region,
$$
\V^{+, n+1}_{\kappa, 2\tau}(\pb_{n+1}) = \selfnonekappa \; .
$$
Thus the contribution is
\begin{equation}\label{goodup}
\int\;  d \mu_{n+1} (\pb_{n+1}) \;  \good^{2\tau}(\pb_{n+1}) \frac
{| { A_+ F} ( p _1, \cdots,  p _{n+1})|^2} {\lambda+
\bomega(\pb_{n+1})+ \selfnonekappa } \; .
\end{equation}
Since $\V^{+, n+1}_{\kappa, 2\tau} = 0$ in the bad region, the contribution from
this region is
\begin{equation}\label{badup}
\int\;  d \mu_{n+1} (\pb_{n+1}) \;  \bad^{2\tau}(\pb_{n+1}) \frac
{|{ A_+ F} ( p _1, \cdots,  p _{n+1})|^2} {\lambda+
\bomega(\pb_{n+1}) }
\end{equation}

\bigskip
\subsection{Decomposition into Diagonal and Off-Diagonal Terms}

Denote by $\Th$ the function
\begin{equation}\label{Th}
\Th(\pb_{n+1})
= [\lambda+ \bomega(\pb_{n+1})+ \selfnonekappa]^{-1}
\end{equation}
The contribution from the good region  can
be decomposed  into diagonal and off-diagonal terms:
\begin{align}\label{def:diagoff}
& \int\;  d \mu_{n+1} (\pb_{n+1}) \;  \good^{2\tau}(\pb_{n+1})
{\Thnone} {| { A_+ F} ( p _1, \cdots,  p _{n+1})|^2}
\no \\ & =
\frac {n(n+1)}  2 \l \, F, \; \diagtwotkn F \, \r   \no \\
& +  {  n(n-1)} (n+1)\l \, F, \; \offonetwotkn  F\, \r+
 \frac {  n(n-1)(n-2)(n+1)}  4 \l \, F, \;  \offtwotwotkn F\, \r
\end{align}
where
\begin{align}
\l \, F, \; \diagtwotkn F \, \r  & =   \int d \mu_{n+1} (\pb_{n+1}) \;
 {\Thnone} {|e^{i r_n}- e^{-i
r_{n+1}}|^2 }   \no \\
& \qquad \quad  \times {\good^{2\tau}}(\pb_{n+1})
\left | \, {   { F}}(p_1,
\cdots,  p_{n-1}, p_n+p_{n+1})
\, \right |^2  \label{def:diag}\\
\l \, F, \; \offonetwotkn F \, \r & = \frac 12
 \int d \mu_{n+1} (\pb_{n+1}) \; {\good^{2\tau}}(\pb_{n+1}) {\Thnone} {
(e^{i r_1}- e^{ - i r_{n+1}})(e^{i r_2}- e^{ - i r_{n+1}}) } \no\\
& \qquad \times \left [ \overline{   { F}}(p_1+p_{n+1}, p_2
\cdots,  p_{n})   { F}(p_1, p_2+p_{n+1},  \cdots,  p_{n})
 + c.c.\right ]   \label{def:off1} \\
\l \, F, \;  \offtwotwotkn \, F \r & =  \frac 12 \int d \mu_{n+1} (\pb_{n+1}) \;
{\good^{2\tau}}(\pb_{n+1})  {\Thnone} { (e^{i r_1}- e^{-i
r_{2}})(e^{i r_3}- e^{-i r_{4}}) }  \no \\
& \qquad \times \left [ \overline{   { F}}(p_1+p_{2}, p_3 \cdots,
p_{n+1})   { F}(p_1, p_2, p_3+p_{4},  \cdots,  p_{n+1})
 + c.c.\right ]  \label{def:off2}
\end{align}

To check the combinatorics, we notice that the total number of terms
are
$$
  \Big (\frac {n (n+1)} 2 \Big )\Big [ 1 +
 2(n-1)+ \Big ( \frac {(n-1)(n-2)} 2 \Big )
 \Big ] =  \Big (\frac {n (n+1)} 2 \Big )^2
$$
the same as the total number of terms in $(AF)^2$.
The factors are obtained in the following way. Notice that
in the formula of $(AF)^2$ we have to choose two indices.
We first  fix the
special two indices in one $F$ to be, say, $(1,2)$.
This gives a factor $n(n+1)/2$. There is only one
choice for the second index to be $(1,2)$ and this gives the first factor
for the diagonal term. There are  $2(n-1)$ choices to
have either $1$ or $2$ and $(n-1)(n-2)/2$ choices to have neither
$1$ nor $2$. These give the last two factors.

Notice  that by the Schwarz inequality, the off-diagonal term is bounded by
the diagonal term.  For the purposes of upper bound we only have to
estimate the diagonal term. Since the number of the off-diagonal terms
are bigger than the diagonal terms by a factor of order $n^2$, we have
the upper bound
\begin{align}\label{up2}
& \int\;  d \mu_{n+1} (\pb_{n+1}) \;  \good^{2\tau}(\pb_{n+1}) {\Thnone}
{| { A_+ F} ( p _1, \cdots,  p _{n+1})|^2}
\no \\ & \le
C n^4 \l \, F, \; \diagtwotkn F \, \r
\end{align}

\bigskip

\subsection{Preliminary Remarks}

Notice in the expression for $\diagtwotkn$ we can integrate the variables
$p_{n}- p_{n+1}$. So we make the change of variables
and define some notations:
\begin{equation}\label{change of variable}
 u_+= p_{n}+ p_{n+1}, \quad u_-= p_{n}- p_{n+1}, \qquad \sqrt 2 x =
r_{n}- r_{n+1},
 \quad \sqrt 2 y =  s_{n}- s_{n+1}
\end{equation}

Suppose at least one of
$|r_n|, |r_{n+1}|, |s_n|, |s_{n+1}|$ is not near $0$ or $\pi$, say
$$
\pi/100 \le |r_n| < 99 \pi/100
$$
Then
we have
$$
\bomega(\pb_{n+1})\ge \omega(r_n) \ge C
$$
for some constant. Therefore, we can bound the kernel
$\Thnone \le C^{-1} $ and
\begin{equation}\label{up1.1}
|e^{i r_n}- e^{-i
r_{n+1}}| = |e^{i (r_n+ r_{n+1})}-1|\le C \sqrt { \omega(r_n+ r_{n+1})}
\end{equation}
After integrating $p_{n}- p_{n+1}$,
we change the variable
$u_+ = p_{n}+ p_{n+1}$ to $p_n$. Recall the normalization difference $((n+1)!)^{-1}$ and $ (n!)^{-1}$ for
$d \mu_{n+1}$ and $d \mu_{n}$.
Thus we have
\begin{align}\label{preli}
& \int d \mu_{n+1} (\pb_{n+1}) \;\{ \pi/100 \le |r_n| < 99 \pi/100\}
 \no \\
 & \qquad {\Thnone} {|e^{i r_n}- e^{-i
r_{n+1}}|^2 }
\left | \, F(p_1,
\cdots,  p_{n-1}, p_n+p_{n+1})
\, \right |^2 \no \\
\le&  C n^{-1} \int d \mu_{n} (\pb_{n}) \;\omega ( r_n)
\left | \, {   { F}}( \pb_n)\, \right |^2
\end{align}
Since we are interested only in terms diverge as $\lambda \to \infty$,
this term is negligible.
Therefore, we shall assume that
\begin{equation}\label{precut1}
|r_n|, |r_{n+1}|, |s_n|, |s_{n+1}| \in [0, \pi/100] \cup [99 \pi/100, \pi]
\end{equation}

We now divide the integration region according to $|r_n|, |r_{n+1}|,
|s_n|, |s_{n+1}|$ in $[0, \pi/100]$ or $[99 \pi/100, \pi]$. There are sixteen
disjoint regions and the final results are obtained by
adding together the estimates from these sixteen
disjoint regions. For simplicity, we shall consider only the region that all
these variables are in the interval $[0, \pi/100]$. The estimates
in all other regions are the same. For example, suppose that
$r_{n+1} \in [99 \pi/100, \pi]$ and the other three variables belong to
$[0, \pi/100]$.
Let  $  p_{n+1}= (\pi, 0)+ \tilde p_{n+1}$ and define
$$
G(\pb_{n},\tilde p_{n+1})
 = F(\pb_{n+1})
$$
Now we have $|\tilde r_{n+1}|, |\tilde  s_{n+1}| \in [0, \pi/100]$
and we can perform estimation on $G$ instead on $F$. Notice that the factor
$e^{i r_n}- e^{-ir_{n+1}}$ may change by a factor of modular one.
But we shall always take its absolute value in the estimates
so that this factor disappears.

Therefore, we now
assume the following generality assumption
\begin{equation}\label{g1}
G_I:\qquad |r_n|, |r_{n+1}|, |s_n|, |s_{n+1}| \in [0, \pi/100]
\end{equation}
This argument applies  to all terms for the rest of this paper
and we shall from now on consider only this case.
The indices $n, n+1$ are the two indices appear
in $F(p_1,\cdots,  p_{n-1}, p_n+p_{n+1})$; they may change depending on
the variables we use in the future.
Notice in this region, we have
\begin{equation}\label{low21}
 \omega(p_{j}) \sim p_j^2 \; , \; \; j=n , n+1,  \quad
 \omega(p_{n}\pm p_{n+1})\sim (p_{n}\pm p_{n+1})^2
\end{equation}
Since we concern only the order of magnitude, {\it for
the rest of the proof for Theorem \ref{mainest} in sections 4-6, we shall
replace $\omega(p)$ by $p^2$ whenever it is more convenient.}

\subsection{Upper Bound of the Diagonal Term: the Good Region}

The following Lemma is the  main estimate on the
diagonal term in the good region.

\begin{lemma}\label{le:goodup}
\begin{equation}\label{up4}
 \l \, F, \; \diagtwotkn F \, \r   \le  \frac C { (n+1)}\int d \mu_n(\pb_n) \;
\rselfnkappatilde
\left | F (\pb_{n})\right |^2
\end{equation}
\end{lemma}

Recall the change of variables \eqref{change of variable}.
We can bound the diagonal term from above  as
\begin{align*}
& \l \, F, \; \diagtwotkn F \, \r  \\
& \le \frac C { (n+1)!}\int_{\sum_{j=1}^{n-1} p_j+ u_+=0}
\prod_{j=1}^{n-1}d p_j\; \int du_+   \; \omega(e_1 \cdot u_+)
\good^{2\tau}(\pb_{n+1})
\left | {   { F}}(p_1,  \cdots,  p_{n-1}, u_+)\right |^2 \\
& \times \int_{-\pi/10}^{\pi/10} \int_{-\pi/10}^{\pi/10} d x d y  \; \Big [\lambda +
a^2 + b^2 + x^2 + y^2 +
\big ( a^2+ x^2 \big ) \big |\log \big (\lambda+ a^2+ b^2+ x^2+ y^2\big ) \big|^{\kappa}
 \Big ]^{-1}
\end{align*}
where
$$
b^2= \bomega(\sb_{n-1}) + \omega(e_2\cdot u_+), \quad a^2 = \bomega(\rb_{n-1}) +
\omega(e_1\cdot u_+) .
$$

Clearly, we have
$$
\good^{2\tau}(\pb_{n+1}) \subset
\big \{ x^2+y^2
\le C |\log \lambda|^{-4\tau} \big \}
\big \{ a^2+ b^2
\le C |\log \lambda|^{-4\tau} \big \}
$$
We now replace
$u_+$ by $p_n$.
Recall the normalization difference $((n+1)!)^{-1}$ and $ (n!)^{-1}$ for
$d \mu_{n+1}$ and $d \mu_{n}$.
Thus we have the upper bound
\begin{align*}
 \l \, F, \; \diagtwotkn F \, \r   \le
 & \frac C { (n+1)}\int d \mu_n(\pb_n) \; \omega(r_{n})
\left | F (\pb_{n})\right |^2 \big \{ a^2+ b^2
\le C |\log \lambda|^{-4\tau} \big \}\\
& \times \int \int dx d y  \;\big \{ x^2+y^2
\le C |\log \lambda|^{-4\tau} \big \}\\
& \times  \Big [ \lambda+ b^2 +
 y^2 +  (a^2+ x^2) \big \{ 1+  |\log (\lambda+ a^2+ b^2+x^2+ y^2)|^{\kappa} \big \}
 \Big ]^{-1}
\end{align*}
where
$$
b^2= \bomega(\sb_{n}) , \quad a^2 = \bomega(\rb_{n})
$$
We need the following Lemma which will be used in several  places later on.

\bigskip

\begin{lemma}\label{le:up}
Let $\tau > 1$ and
\begin{align*}
K_\kappa^\tau (a, b)=&
 \int \int dx d y  \, \big \{ x^2+y^2 \le |\log \lambda|^{-2\tau} \big \}
\\ &   \Big [ \lambda+ b^2 +
 y^2 +  (a^2+ x^2) \big \{ 1+  |\log (\lambda+ a^2+ b^2+x^2+ y^2)|^{\kappa} \big \}
 \Big ]^{-1}
\end{align*}

Suppose that
\begin{equation}\label{abup}
a^2+ b^2 \le \logl^{-2\tau}
\end{equation}
Then for $0 \le \kappa \le 1$ we have
\begin{equation}\label{eq:upest}
K_\kappa^\tau (a, b)\le  C \big |\,  \log (\lambda+ a^2+ b^2) \, \big |^{1-\kappa/2}
\end{equation}
On the other hand, if
\begin{equation}\label{abup2}
a^2+ b^2 \le \logl^{-4\tau},
\end{equation}
we have the lower bound
\begin{equation}\label{eq:lowest}
K_\kappa^\tau (a, b) \ge C^{-1}  \big |\,  \log (\lambda+ a^2+ b^2)
\, \big |^{1-\kappa/2}
\end{equation}
We also have the trivial bound
\begin{equation}\label{trivialup}
\int_{-\pi}^\pi \int_{-\pi}^\pi d x d y \Big [ \lambda+  a^2+ b^2
 + y^2    + x^2 \Big ]^{-1}
 \le C \big |\,  \log  (\lambda+ a^2+ b^2)
\, \big |
\end{equation}
\end{lemma}

\myproof 
Clearly the trivial bound can be checked easily. We now prove the rest.
Fix a constant $m$, $1 < m < \tau$.
Let
$$
 \goodone (x, y) = \Big \{\, (x, y):
  |x| \le   |\log \lambda|^{m}  |y| \le
 |\log \lambda|^{2m}  |x|\; \Big \}
$$
and $ \badone$ be its complement.

In the region
$ \badone$ we drop $ (a^2+ x^2)
\big \{ 1+  |\log (\lambda+ a^2+ b^2+ y^2)|^{\kappa} \} $ to
have an upper bound.
The angle integration of $x, y$
gives a factor $\logl^{-m}$. Thus
the contribution from this region is bounded by
$$
 C  \logl^{-m+1} \le C   \,
$$

\medskip

In the region $\goodone$ we have
\begin{align*}
& \log (\lambda+ a^2+ b^2+ y^2)^{-1} -   \log ( 1 + \logl^{2 m})\\
 \le & \log (\lambda+ a^2+ b^2+ x^2+ y^2)^{-1} \\
 \le &
\log (\lambda+ a^2+ b^2+ y^2)^{-1}
\end{align*}
By assumptions \eqref{abup} or \eqref{abup2},
$ a^2+ b^2+ x^2+y^2 \le 2 |\log \lambda|^{-2\tau} $.
Thus for $\tau > m$, we have
$$
\log (\lambda+ a^2+ b^2+ y^2)^{-1} -   \log ( 1 + \logl^{2 m})
\ge C \log (\lambda+ a^2+ b^2+ y^2)^{-1}
$$
for some constant depends on ${\tau, m}$.
Therefore, we have
\begin{equation}
 C \log (\lambda+ a^2+ b^2+ y^2)^{-1}
 \le  \log (\lambda+ a^2+ b^2+ x^2+ y^2)^{-1}
\le \log (\lambda+ a^2+ b^2+ y^2)^{-1} \; .
\end{equation}

\medskip \noindent
{\it Upper Bound}: \quad  We now replace $\log (\lambda+ a^2+ b^2+ x^2+ y^2)^{-1}$
by $ C  \log (\lambda+ a^2+ b^2+  y^2)^{-1}$
and drop $a^2 \{ 1+ |\log (\lambda+ a^2+ b^2+x^2+  y^2)|^{\kappa}\}$
to have an upper bound
for $K_\kappa^\tau$. Thus we can bound $K_\kappa^\tau (a, b) $ by
\begin{align*}
K_\kappa^\tau (a, b) \le &
C  \int \int dx d y  \, \big \{ x^2+y^2 \le |\log \lambda|^{-2\tau} \big \}
\\ &   \Big [ \lambda+ b^2 +
 y^2 +  a^2+ x^2  |\log (\lambda+ a^2+ b^2+ y^2)|^{\kappa} \big \}
 \Big ]^{-1}
\end{align*}

Change the variable by
$$
z = x |\log (\lambda+ a^2+ b^2+ y^2)|^{\kappa/2}
$$
Hence
$$
z^2\le x^2 |\log (\lambda+ a^2+ b^2)|^{\kappa}
$$
Thus for $x, y$ in the integration region we have
$$
 {z^2}  +y^2 \le |\log \lambda|^{-2\tau}
|\log (\lambda+ a^2+ b^2)|^{\kappa}\le C
$$
We can bound $K_\kappa^\tau (a, b) $ by
\begin{align*}
K_\kappa^\tau (a, b) \le &
C  \int \int d  z d y  \,
\big \{ {z^2}  +y^2 \le C \big \} \\
&  \Big [ \lambda+  a^2+ b^2
 + y^2  + z^2  \Big ]^{-1}
|\log (\lambda+ a^2+ b^2+ y^2)|^{-\kappa/2}
\end{align*}
Denote by
\begin{equation}
\rho^2 = z^2+ y^2
\end{equation}
Since
$$
\log (\lambda+ a^2+ b^2+ y^2)^{-1} \ge \log (\lambda+ a^2+ b^2+ \rho^2)^{-1},
$$
we can bound the integration by
$$
C \int_0^C  d \rho^2  \big (\lambda+   a^2+ b^2 + \rho^2 \big )^{-1}
 |\log (\lambda+ a^2+ b^2+
\rho^2)|^{-\kappa/2} \le C   |\log (\lambda+ a^2+ b^2)|^{1-\kappa/2}
$$
This proves the upper bound.

\medskip \noindent
{\it Lower Bound}: \quad
We now replace $\log (\lambda+ a^2+ b^2+ x^2+ y^2)^{-1}$
by $  \log (\lambda+ a^2+ b^2+  y^2)^{-1}$
to have a lower bound for $K_\kappa^\tau$.
We  change the variable to the same $z$ and $\rho$ as in the upper bound.
We now restrict the angle $\theta(z, y)$ of the two dimensional vector
$(z, y)$ to be between $\pi/3$ and $2 \pi/3$, i.e.,
\begin{equation}\label{low26}
\pi/3 \le \theta(z, y)  \le 2 \pi/3
\end{equation}
In this region, $|z|\sim |y|\sim \rho$.
Denote by  $q = \rho^2 + a^2+ b^2$. We further restrict the integration to
\begin{equation}\label{low27}
{\mathcal W} = \big \{ \,2 ( a^2 |\log (\lambda+ a^2+ b^2)|^{\kappa}+  b^2) \le q \le |\log
\lambda|^{-2\tau}/2 \, \big \}
\end{equation}
From the last restriction, we also have $ \rho^2 \le |\log
\lambda|^{-2\tau}/2$. Since $a^2+b^2$ satisfies \eqref{abup2}
and $|x| \le |z|$,
the condition $ x^2+y^2 \le |\log \lambda|^{-2\tau} $ is satisfied.

The integral  is thus bounded below by
\begin{align*}
  \int d z d y  & \; \{ \pi/3 \le \theta(z, y)  \le 2 \pi/3 \}
  {\mathcal W} (q)    |\log
(\lambda+ a^2+b^2+ \rho^2)|^{-\kappa/2}  \\
&\Big [ \lambda+  b^2 + \rho^2 +  a^2 |\log
(\lambda+ a^2+b^2+ \rho^2)|^{\kappa} \Big ]^{-1}
\end{align*}

From the restriction on $q$, we have
$$
\rho^2  \ge   a^2|\log (\lambda+ a^2+ b^2)|^{\kappa}
\ge a^2 |\log (\lambda+ a^2+ b^2+ \rho^2)|^{\kappa}
$$
Thus
$$
\Big [ \lambda+  b^2 + \rho^2 +  a^2 |\log
(\lambda+ a^2+b^2+ \rho^2)|^{\kappa} \Big ]^{-1} \ge (1/2) (\lambda+ a^2+ b^2 + \rho^2
)^{-1}
$$
The angle integration produces just some constant factor. Thus the integral is bigger than
\begin{eqnarray}\label{low1.1}
 &&
 C \int_{2 (a^2 | \log a^2|^\kappa+  b^2)}^{|\log \lambda|^{-2\tau}/2}
  d q\,   (\lambda+ q)^{-1} |\log (\lambda+ q) |^{-\kappa/2} \no \\
\ge &&  C \Big [ \big | \log ( \lambda+  2 a^2 | \log a^2|^\kappa+  2 b^2 ) \big
|^{1-\kappa/2} - \big [ 2\tau \log |\log \lambda| + \log 2 \big
]^{1-\kappa/2} \Big ]
\end{eqnarray}
Since $a^2+ b^2 \le |\log \lambda|^{-4 \tau} $
we have
\begin{equation}\label{low8}
\log  (\lambda+  2 a^2 | \log a^2|^\kappa+  2 b^2)^{-1} \ge
(7\tau/  2)   |\log   \log \lambda|
\end{equation}
Therefore, we have the bound
\begin{align*}
& \big | \log (\lambda+  2 a^2 | \log a^2|^\kappa+  2 b^2) \big |^{1-\kappa/2} -
\big [ 2\tau \log |\log \lambda| + \log 2 \big ]^{1-\kappa/2} \\
& \ge
(1/20) \big | \log (\lambda+a^2 +  b^2)   |^{1-\kappa/2}
\end{align*}
We have thus proved the lower bound.
\myendproof

\bigskip

From this Lemma, we have proved Lemma \ref{le:goodup}
concerning the estimate in the good region.
Observe that the main contribution of the $p_n-p_{n+1}$ integration comes from
the region $|p_n-p_{n+1}|  \gg  |p_n+p_{n+1}|  + \bomega(\pb_{n-1}) $.
In fact, we have the following
Lemma.

\bigskip

\begin{lemma}\label{error}
For any  $m> 0$ there is a constant $C_m$ such that
\begin{align}
&  \int d \mu_{n+1} (\pb_{n+1}) \; \Big \{
 |p_n-p_{n+1}|^2    \le \logl^{2m} \big [\,
 |p_n+p_{n+1}|^2  + \bomega(\pb_{n-1})\, \big ] \Big \}\no \\
& \qquad \times   \frac {|e^{i r_n}- e^{-i r_{n+1}}|^2 }
 {  \lambda+ \bomega(\pb_{n+1}) }
\left | \, F (p_1,  \cdots,  p_{n-1}, p_n+p_{n+1}) \, \right |^2   \no \\
&  \le
 C_m n^{-1}  \, \loggl
 \int d \mu_{n} (\pb_{n}) \;  \omega(r_{n})
 \left | \, {   { F}}(\pb_n)
\, \right |^2 \label{badup5}
\end{align}
\end{lemma}

\myproof  We have the bound
\begin{align*}
& \int d(p_n-p_{n+1})\;
\frac { |e^{i r_n}- e^{-i r_{n+1}}|^2 }{ \lambda+ \bomega(\pb_{n+1})  } \\
& \qquad \quad \times \big \{ |p_n-p_{n+1}|^2    \le \logl^{2m} \big [\,
 |p_n+p_{n+1}|^2  + \bomega(\pb_{n-1})\, \big ]
 \big \} \\
&
\le   \log \Big ( \frac {\lambda+
(1+\logl^{2m}) \big [  |p_n+p_{n+1}|^2  + \bomega(\pb_{n-1})\big ] }
{  \lambda+  |p_n+p_{n+1}|^2  + \bomega(\pb_{n-1})} \Big )
\le C_m \loggl
\end{align*}
Changing the variable $p_n+p_{n+1} \to p_n$, we have proved the Lemma.
\myendproof

Therefore, with a price of the term on the right side of \eqref{badup5}
we can assume the following general assumption (II)
\begin{equation}\label{g2}
G_{II}: \qquad  |p_n-p_{n+1}|^2    \ge \logl^{2m} \big [\,
 |p_n+p_{n+1}|^2  + \bomega(\pb_{n-1})\, \big ] \; .
\end{equation}
Under the assumptions  \eqref{mbound} \eqref{nbound},
the term on the right side of \eqref{badup5} is much smaller than the accuracy
we need for Theorem \ref{mainest}.
Therefore this condition will be imposed for the rest of the paper.

\bigskip
\subsection{Upper Bound of the Diagonal Term: the Bad Region}


The contribution from the bad region  can
be decomposed  into diagonal and off-diagonal terms. Again, we shall use
the Schwarz inequality to bound the off-diagonal terms by the diagonal terms.
Therefore, we have the bound
\begin{align*}
& \int\;  d \mu_{n+1} (\pb_{n+1}) \;  \bad^{2\tau}(\pb_{n+1})
\frac {| { A_+ F} ( p _1, \cdots,  p _{n+1})|^2}
{\lambda+ \bomega(\pb_{n+1}) } \\
& \le 2 n^4 \int d \mu_{n+1} (\pb_{n+1}) \;
\bad^{2\tau}(\pb_{n+1})\frac {|e^{i r_n}- e^{-i r_{n+1}}|^2
} {  \lambda+\bomega(\pb_{n+1}) }  \left | \, {   { F}}(p_1,
\cdots,  p_{n-1}, p_n+p_{n+1}) \, \right |^2  
\end{align*}
Again the variable  $ p_n-p_{n+1}$ does not appear in $F$
and we can perform the integration.

We subdivide $\bad^{2\tau}(\pb_{n+1})$ into
$$
\bad^{2\tau}(\pb_{n+1})\bad_{n}^{4\tau}(\pb_{n-1},
p_n+p_{n+1})\;  \cup  \;  \bad^{2\tau}(\pb_{n+1})\good^{4\tau}(\pb_{n-1},
p_n+p_{n+1}) \; .
$$
In the first case, we  drop the characteristic
function $\bad^{2\tau}(\pb_{n+1})$ to have an upper bound.
We now use the trivial bound \eqref{trivialup} to estimate the integration
of the variable $ p_n-p_{n+1}$ by
\begin{align}\label{badup1}
& 2 n^4 \int d \mu_{n+1} (\pb_{n+1}) \;
\bad^{4\tau}(\pb_{n-1},
p_n+p_{n+1})  \no \\
& \qquad \quad \times \frac {|e^{i r_n}- e^{-i r_{n+1}}|^2 } {  \lambda+
\bomega(\pb_{n+1}) }
 \left | \, F(p_1,  \cdots,  p_{n-1}, p_n+p_{n+1})
\, \right |^2   \no \\
&  \le
 C n^{3}  \,
 \int d \mu_n (\pb_{n}) \;
\bad^{4\tau}(\pb_{n}) \rselfnone
 \left | \, F(p_1,  \cdots,   p_n)
\, \right |^2
\no \\
&  \le
 C n^{3}  \loggl \,
 \int d \mu_n (\pb_{n}) \; \omega(r_n)
 \left | \, F(p_1,  \cdots,   p_n)
\, \right |^2
\end{align}
Here we have used the change of the normalization between
$d \mu_{n+1}$ and $d \mu_{n}$.

\bigskip

We now estimate the region $\bad^{2\tau}(\pb_{n+1})\good^{4\tau}(\pb_{n-1},
p_n+p_{n+1})$  which is the transition from the bad set to good set.
In this region,
$$
|p_n-p_{n+1}|^2 \ge C |\log \lambda|^{-4\tau}
$$
The contribution  is bounded by
\begin{align}
& 2 n^4 \int d \mu_{n+1} (\pb_{n+1}) \; \Big
\{|p_n-p_{n+1}|^2\ge C |\log
\lambda|^{-4\tau} \Big \}  \no \\
& \qquad \times  \frac {|e^{i r_n}- e^{-i r_{n+1}}|^2 }
{  \lambda+ \bomega(\pb_{n+1}) }
 \left | \, {   { F}}(p_1,  \cdots,  p_{n-1}, p_n+p_{n+1})
\, \right |^2   \no \\
&  \le
 C n^{3} |\log \log \lambda|  \,
 \int d \mu_{n} (\pb_{n}) \;  \omega(r_{n})
 \left | \, {   { F}}(\pb_n)
\, \right |^2 \label{badup2}
\end{align}

Combining the estimates  \eqref{badup1} and \eqref{badup2},
we can bound the contribution from the bad region by
\begin{align*}
& \int\;  d \mu_{n+1} (\pb_{n+1}) \;  \bad^{2\tau}(\pb_{n+1})
\frac {| { A_+ F} ( p _1, \cdots,  p _{n+1})|^2}
{\lambda+ \bomega(\pb_{n+1}) } \\
& \le C n^{3} |\log \log \lambda|  \,
 \int d \mu_{n} (\pb_{n}) \;  \omega(r_{n})
 \left | \, {   { F}}(\pb_n)
\, \right |^2
\end{align*}
Together with the estimate on the good region, Lemma \ref{le:goodup},
we can bound the right side of \eqref{up12} by
\begin{align*}
& \int\;  d \mu_{n+1} (\pb_{n+1}) \;  \frac {|
{ A_+ F} ( p _1, \cdots,  p _{n+1})|^2} {\lambda+ \bomega(\pb_{n+1})+
\V^{+, n+1}_{\kappa, 2\tau}(\pb_{n+1}) } \; \\
\le &   C n^3 \int d \mu_n(\pb_n) \;
\rselfnkappatilde
\left | F (\pb_{n})\right |^2 \no \\
&  + C n^{3} |\log \log \lambda|  \,
 \int d \mu_n (\pb_{n}) \;  \omega(r_{n})
 \left | \, F(\pb_n)
\, \right |^2
\end{align*}
Under the condition \eqref{nbound}, it is easy to check that
for symmetric function $F$ the right side
of the last equation is bounded above by $|\log \log \lambda|^2
\U^n_{\tilde  \kappa,  \tau}  (\pb_n)$. This proves the upper
bound for Theorem \ref{mainest}.

\section{Lower Bound: The Diagonal terms}

By definition, we have
\begin{align}
& \l F A^* (\lambda- S_{n+1}+ \gamma^{-1} \U^{n+1}_{\kappa, \tau})^{-1} A  F \r  \no \\
& = \int\;  d \mu_{n+1} (\pb_{n+1}) \;  \frac {| { A_+ F} ( p
_1, \cdots,  p _{n+1})|^2}
{\lambda+ \bomega(\pb_{n+1})
+ \gamma^{-1}  \U^{n+1}_{\kappa, \tau}(\pb_{n+1}) }
\label{low0}
\end{align}
Since $\gamma \le 1$ and $\lambda+ \bomega\ge 0$, 
the integral is bigger than
\begin{equation}
\gamma \int\;  d \mu_{n+1} (\pb_{n+1}) \;  \frac {|   { A_+ F}
(p_1, \cdots,  p _{n+1})|^2}
{\lambda+ \bomega(\pb_{n+1})+   \U^{n+1}_{\kappa, \tau}(\pb_{n+1}) }  \no
\label{low1}
\end{equation}

Divide the integral into $\pb_{n+1} \in \good^{\tau}$ and
$\pb_{n+1} \in \bad^{\tau}$.  In the bad set
$\bad^{\tau}$, we bound the integral in this region from
below by zero.
In the good set, we have
$$
\U^{n+1}_{\kappa,  \tau}(\pb_{n+1}) = \selfnonekappa  , \qquad
\pb_{n+1} \in
\good^{\tau}
$$
Thus
\begin{eqnarray} \label{low3}
& &\l F A^* (\lambda- S_{n+1}+   \U^{n+1}_{\kappa, \tau})^{-1} A  F \r  \no \\
&  & \ge
\int\;  d \mu_{n+1} (\pb_{n+1}) \; {\good^{\tau}}(\pb_{n+1})
\frac {|   { A_+ F} ( p _1, \cdots,  p _{n+1})|^2} {\lambda+
\bomega(\pb_{n+1})+
\U^{n+1}_{\kappa, \tau}(\pb_{n+1}) }  \no \\
& & \ge  \int\;  d \mu_{n+1} (\pb_{n+1}) \;
{\good^{\tau}}(\pb_{n+1}) \Th(\pb_{n+1}) |   { A_+ F} ( p _1,
\cdots,  p _{n+1})|^2
\end{eqnarray}
where $\Th(\pb_{n+1})$ is defined in \eqref{Th}.
We now decompose the last term into diagonal and off-diagonal terms:
\begin{align}
 &\frac {n(n+1)}  2 \l \, F, \; \diagtkn F \, \r   \no \\
& +  {  n(n-1)} (n+1)\l \, F, \; \offonetkn  F\, \r+
 \frac {  n(n-1)(n-2)(n+1)}  4 \l \, F, \;  \offtwotkn F\, \r
\label{def:diagoff2}
\end{align}
where these operators are defined in \eqref{def:diag}-\eqref{def:off2}.

\subsection{Lower Bound on the Diagonal Terms}

The main estimate on the lower bound of the diagonal term
\eqref{def:diag} is  the
following Lemma.
Define
\begin{equation}\label{Fgood}
F_\good^{ 2 \tau} (\pb_{n}) =   { F}(\pb_{n})
\good^{ 2 \tau} (\pb_{n}), \qquad
F_\bad^{ 2 \tau} (\pb_{n}) =   { F}(\pb_{n})
\bad^{ 2 \tau} (\pb_{n})
\end{equation}

\bigskip

\begin{lemma}\label{le:diaglow}
Recall $\kappa, \tau$ and $n$ satisfy the assumptions
\eqref{mbound} and \eqref{nbound}.
Then the diagonal term  is bounded below by
\begin{equation}
\l \, F, \; \diagtkn\,  F \, \r \ge  \; C  n^{-1} \, \l \,
F_\good^{ 2 \tau}  , \; \rselfnkappatilde \,  F_\good^{ 2 \tau}  \, \r
\end{equation}
\end{lemma}
\bigskip

\myproof
Recall the assumptions \eqref{g1}, \eqref{g2} and the
change the variables
\begin{align}\label{change of variable2}
& u_+= p_{n}+ p_{n+1}, \quad u_-= p_{n}- p_{n+1}, \qquad \sqrt 2 x =
r_{n}- r_{n+1},
 \quad \sqrt 2 y =  s_{n}- s_{n+1} \no \\
&
b^2 =  { \omega(e_2 \cdot u_+)} + \bomega(\sb_{n-1}), \quad a^2 =
{ \omega(e_1 \cdot u_+)} + \bomega(\rb_{n-1})
\end{align}
Thus we can bound the diagonal term from below by
\begin{align*}
& \l \, F, \; \diagtkn F \, \r  \\
& \ge \frac C {(n+1)!}\int_{\sum_{j=1}^{n-1} p_j+ u_+=0}
\prod_{j=1}^{n-1}d p_j\; \int du_+   \; \omega(e_1 \cdot u_+)
\good^\tau(\pb_{n+1})
\left | F(p_1,  \cdots,  p_{n-1}, u_+)\right |^2 \\
& \times \int \int dx d y  \;  \Big [\lambda +  a^2 + b^2 + x^2 + y^2 +
(a^2+ x^2) |\log (\lambda+a^2+ b^2+ x^2+y^2)|^{\kappa} \Big ]^{-1}
\end{align*}
We now impose the condition $x^2 + y^2 \le \logl^{-2 \tau} /2$
to have a lower bound.
Since
$$
\good^{2\tau}( p_1,  \cdots,  p_{n-1}, u_+)\, \big \{\,
x^2 + y^2 \le \logl^{-2 \tau} /2 \, \big \}
\subset \good^\tau(\pb_{n+1}),
$$
we can replace $\good^\tau(\pb_{n+1})$ by $\big \{\,
x^2 + y^2 \le \logl^{-2 \tau} /2 \, \big \}$
and  $F$ by $F_\good^{2 \tau}$ to have  a lower bound.
The lemma now
follows from the lower bound of Lemma \ref{le:up}. \myendproof

\bigskip
\section{Off-diagonal terms}
\bigskip

Our goal in this section is to prove the following estimate
on the off-diagonal terms.

\begin{lemma}\label{le:off1}
Recall that $\kappa,  \tau$ and $n$ satisfy the assumptions
\eqref{mbound} and \eqref{nbound}.
The first and second off-diagonal terms are bounded by
\begin{align}
&  \big | \l \, F, \; \offonetkn F \, \r \big |
+ \big | \l \, F, \; \offtwotkn F \, \r \big | \no \\
 \le &
C n^{-1} |\log \log
\lambda |^{1+ 1/2} \int d \mu_{n} (\pb_{n})\;
\omega(r_{n})  \, \big | {F}_\bad^{ 2 \tau} (\pb_{n})
\big |^2  \no \\
& + C  n^{-5} \,
  \int d \mu_n(\pb_n) \;
 \omega(r_{n}) |\log (\lambda+ \bomega(\pb_n) |^{1-\kappa/2}
\left | F_\good^{ 2 \tau} (\pb_{n})\right |^2
\end{align}
\end{lemma}

\bigskip
\myproof
The first off-diagonal term is bounded by
\begin{align*}
& \Big | \l \, F, \; \offonetkn F \, \r \Big |  \\
& \le
C  \int d \mu_{n+1} (\pb_{n+1}) \;  {\good^{\tau}}(\pb_{n+1}) \Thnone \big | {
(e^{i r_1}- e^{ - i r_{n+1}})(e^{i r_2}- e^{ - i r_{n+1}}) } \big | \no\\
& \qquad \times \Big | {   { F}}(p_1+p_{n+1}, p_2
\cdots,  p_{n})
   { F}(p_1, p_2+p_{n+1},  \cdots,  p_{n})
\Big |   
\end{align*}
By definition $
F = { F}_\good^{ 2 \tau}  + {
F}_\bad^{ 2 \tau} $.
Thus the last term  is equal to
\begin{align*}
& C  \int d \mu_{n+1} (\pb_{n+1}) \;  {\good^{\tau}}(\pb_{n+1}) \Thnone \big | {
(e^{i r_1}- e^{ - i r_{n+1}})(e^{i r_2}- e^{ - i r_{n+1}}) } \big | \no\\
& \qquad \times \Big | \big (  { F}_\good^{ 2 \tau}  + {
F}_\bad^{ 2 \tau} \big ) (p_1+p_{n+1},   p_2,  \cdots,  p_{n}) \\
& \qquad \times\big
(  { F}_\good^{ 2 \tau}  + { F}_\bad^{ 2 \tau} \big )
(p_2+p_{n+1},   p_1, p_3,  \cdots,  p_{n}) \Big |
\end{align*}

From the Schwarz inequality, the cross term
is bounded by
\begin{align}\label{low6}
& C  \int d \mu_{n+1} (\pb_{n+1}) \;  {\good^{\tau}}(\pb_{n+1}) \Thnone \big | {
(e^{i r_1}- e^{ - i r_{n+1}})(e^{i r_2}- e^{ - i r_{n+1}}) } \big | \no\\
& \qquad \times \Big |  { { F}_\good^{ 2 \tau} }
(p_1+p_{n+1},   p_2,  \cdots,  p_{n})
 { F}_\bad^{ 2 \tau}  (p_2+p_{n+1},   p_1, p_3,  \cdots,  p_{n})
\Big | \no \\
& \le C \delta \int d \mu_{n+1} (\pb_{n+1}) \; {\good^{\tau}}(\pb_{n+1})
 \Thnone   { |(e^{i r_1}-
e^{ - i r_{n+1}})|^2 }  \no \\
& \qquad \qquad\qquad \qquad\qquad \times \Big |  { {
F}_\good^{ 2 \tau} }
(p_1+p_{n+1},   p_2,  \cdots,  p_{n}) \Big |^2  \no \\
& + C \delta^{-1} \int d \mu_{n+1} (\pb_{n+1}) \; {\good^{\tau}}(\pb_{n+1})
 \Thnone   { |(e^{i r_2}-e^{ - i r_{n+1}})|^2 }
\no \\
& \qquad \qquad\qquad \qquad\qquad \times \Big |
 { F}_\bad^{ 2 \tau}  (p_2+p_{n+1},   p_1, p_3,  \cdots,  p_{n})
\Big |^2
\end{align}

We first bound the last term. Clearly, in the region
$$
{\good^{\tau}}(\pb_{n+1})
\bad^{ 2 \tau}  \{ p_2+p_{n+1},   p_1, p_3,  \cdots,  p_{n}\}
$$
we have
$$
|p_2-p_{n+1}|^2    \le \logl^{4 \tau} \big [\,
 |p_2+p_{n+1}|^2  + \omega(p_{1})+ \omega(p_{3})+\cdots
 \omega(p_{n})\, \big ]
$$
Thus we can apply Lemma \ref{error}.
Let $\delta = |\log \log \lambda |^{-1/2}$. We can bound the last term
in \eqref{low6} by
\begin{equation}\label{low6.1}
C n^{-1} |\log \log
\lambda |^{1+ 1/2} \int d \mu_{n} (\pb_{n})\;
\omega(r_{n})  \, \big | {F}_\bad^{ 2 \tau} (\pb_{n})
\big |^2
\end{equation}
The first term on the right side of \eqref{low6} can be bounded as
in the section of upper bound. Using Lemma
\ref{le:goodup}, we  bound it by
\begin{equation}\label{low6.2}
C n^{-1} |\log \log \lambda |^{-1/2} \int d \mu_n(\pb_n) \;
\omega(r_{n}) |\log (\lambda+ \pb_n^2) |^{1-\kappa/2}
\left | F_\good^{ 2 \tau} (\pb_{n})\right |^2
\end{equation}

The contribution from the term with $ { F}_\bad^{ 2 \tau}   {
F}_\bad^{ 2 \tau} $ can be estimated similarly. Finally, we
consider the contribution from $ { F}_\good^{ 2 \tau}  {
F}_\good^{ 2 \tau} $. To estimate this term, we need the following
Lemma which  will be proved in the next section.

\begin{lemma}\label{le:off}
Recall that $\kappa, \tau$ and $n$ satisfy the assumptions
\eqref{mbound} and \eqref{nbound}.
Then we have the following two estimates:
\begin{align}
 Q_1= & \int d \mu_{n+1} (\pb_{n+1}) \;  \Th(\pb_{n+1}) \big |  { (e^{i r_1}-
e^{ - i r_{3}})(e^{i r_2}- e^{ - i r_{3}}) }  \big | \no \\
& \qquad \Big |{ { {F}_\good^{ 2 \tau}  }} (p_1+p_{3},   p_2, p_4,
\cdots, p_{n+1}) { {F}_\good^{ 2 \tau}  }( p_2+p_{3},  p_1, p_4,
\cdots,  p_{n+1}) \Big |
\no \\ &
\le C n^{-5} \int d \mu_n (\pb_{n}) \;
\rselfnkappatilde \left |F_\good^{ 2 \tau} (\pb_{n})\right |^2
 \label{off1}
\end{align}
\begin{align}
   Q_2 = & \int d \mu_{n+1} (\pb_{n+1}) \;  \Th(\pb_{n+1}) \big |  { (e^{i r_1}-
e^{ - i r_{2}})(e^{i r_3}- e^{ - i r_{4}}) } \big |\no \\
& \qquad
 \Big |{ { {F}_\good^{ 2 \tau}  }} (p_1+p_{2},   p_3, p_4,
\cdots, p_{n+1}) { {F}_\good^{ 2 \tau}  }(p_3+p_{4},  p_1, p_2, p_5,
\cdots, p_{n+1}) \Big |
\no \\ &
\le C n^{-5} \int d \mu_n (\pb_{n})\;
\rselfnkappatilde \left | {   { {F}_\good^{2 \tau} }}(\pb_n)\right |^2
 \label{off2}
\end{align}

\end{lemma}

We now collect all our efforts. The cross terms are bounded by \eqref{low6.1}
and \eqref{low6.2}. The contribution from $ { F}_\bad^{ 2 \tau}   {
F}_\bad^{ 2 \tau} $ can be estimated similarly. Finally the contribution
from $ { F}_\good^{ 2 \tau}   {
F}_\good^{ 2 \tau} $ is bounded by the last Lemma.
Thus we have proved the estimate on $\offonetkn$ in
Lemma \ref{le:off1}. The estimate on $\offtwotkn$ can be proved
in a similar way by using instead the equation \eqref{off2}.
This proves Lemma \ref{le:off1}.

\subsection{Proof of the Lower Bound}

Recall the condition \eqref{nbound}  on the size of $n$.
Combining the lower bound on the diagonal term in Lemma \ref{le:diaglow}
and the estimate on the off-diagonal terms in Lemma \ref{le:off1},  we have
\begin{align*}
& n^2 \l \, F, \; \diagtkn\,  F \, \r -
n^3 \big | \l \, F, \; \offonetkn F \, \r \big | -
n^4 \big | \l \, F, \; \offtwotkn F \, \r \big |\\
\ge &    C n^2  \int d \mu_n(\pb_n) \;
\rselfnkappatilde
\left | F_\good^{ 2 \tau} (\pb_{n})\right |^2 \\
&  - C n^{4} |\log \log
\lambda |^{1+ 1/2} \int d \mu_{n} (\pb_{n})\;
\omega(r_{n})  \, \big | {F}_\bad^{ 2 \tau} (\pb_{n})
\big |^2  \no \\
& - C \big [\,  n^{-1}+ n^3 \loggl^{-1/2} \, \big] \int d \mu_n (\pb_{n})
\rselfnkappatilde \left | F_\good^{ 2 \tau} (\pb_{n})\right |^2
\end{align*}
The last term can be absorbed into the  first term on the right side
with a change
of constant.  The middle term on the right side gives
the estimate on the bad set.
This proves the lower bound for Theorem \ref{mainest}.

\subsection{Proof of  Lemma \ref{le:off}}

We first bound $Q_1$. Consider the two cases.

\noindent
{\it Case 1}.  Some $p_i, i = 1, 2, 3$ dominates, say, we have
$$
 |p_1| \ge 2 (|p_2| + |p_3|)
$$
Then $
 |p_1-p_3| \le 4 |p_1+p_3|$.
From the Schwarz inequality
\begin{align}\label{o1}
& \int d \mu_{n+1} (\pb_{n+1}) \;  \big \{  |p_1-p_3| \le 4 |p_1+p_3| \big\}
\no \\
& \qquad \times \,  \Thnone \big |{ (e^{i r_1}- e^{ - i r_{3}})(e^{i r_2}- e^{ -
i r_{3}}) } \big |
\no \\ &
\qquad \times \Big |{ { {F}_\good^{ 2 \tau}  }} (p_1+p_{3},   p_2, p_4,
\cdots, p_{n+1}) { {F}_\good^{ 2 \tau}  }(p_2+p_{3},  p_1, p_4,
\cdots,  p_{n+1}) \Big |
\no \\ &
\le \delta^{-1} \int d \mu_{n+1} (\pb_{n+1}) \;
\big \{  |p_1-p_3| \le 4 |p_1+p_3| \big
\}\no \\
& \qquad \times \,  \Thnone { |e^{i r_1}- e^{ - i r_{3}}|^2 }
 \big |{ { {F}_\good^{ 2 \tau}  }} (p_1+p_{3},   p_2, p_4,
\cdots, p_{n+1}) \big |^2
\no \\ &
+ \delta  \int d \mu_{n+1} (\pb_{n+1}) \;  \;
 \Thnone { |e^{i r_2}-
e^{ - i r_{3}}|^2 }
\no \\ &
\qquad \times \big |  { {F}_\good^{ 2 \tau}  }(p_2+p_{3},  p_1, p_4,
\cdots,  p_{n+1}) \big |^2
\end{align}

The last term on the right side of \eqref{o1} can be bounded
using Lemma \ref{le:goodup}.
To estimate the first term,
we  drop $\selfnonekappa$ in $\Thnone$ and integrate $p_1-p_3$.
The integration can be estimated  easily by
\begin{equation}
\int d (p_1-p_3) \;  \big \{  |p_1-p_3| \le 4 |p_1+p_3| \big
\}\;| \Thnone  |
\le C \; .
\end{equation}
We now choose $\delta = n^{-5}$ and use
$$
n^{10} \le \loggl^{1/2} \le C\big | \log (\lambda+ \bomega(\pb_{n})) \big |^{1-\kappa/2}
$$
if $ \bomega(\pb_n) \le |\log \lambda|^{-4\tau}$
and $0\le \kappa \le 1$.
The left side of \eqref{o1} is thus bounded above by
$$
C n^{-5} \int d \mu_n (\pb_{n})  \omega(r_{n})
\big | \log (\lambda+ \bomega(\pb_{n}))
\big |^{1-\kappa/2}
\big |F_\good^{ 2 \tau}  (\pb_{n}) \big |^2
$$
Here we have changed variables so that
the variable of the function
${F}_\good^{ 2 \tau} $ is of the standard form.

\bigskip
\noindent
{\it Case} 2: \quad $|p_1| \sim |p_2| \sim  |p_3|$.

In this case, we have $|p_1-p_3| \le 16 |p_2| $.
Similar arguments prove the same bound in this region.
This proves \eqref{off1}.

\bigskip

We now estimate $Q_2$.
We can assume without loss of generality  that
$$
\omega(p_1-p_2) \le  \omega(p_3-p_4)
$$
Again, we bound it by the Schwarz inequality to have
\begin{align*}
Q_2 \le &   \delta^{-1}   \int d \mu_{n+1}
(\pb_{n+1}) \;  \Thnone {
|e^{i r_1}- e^{ - i r_{2}}|^2 }
\\ &  \qquad  \times
\big \{  \omega(p_1-p_2) \le  \omega(p_3-p_4) \big \}
\big |{ { {F}_\good^{ 2 \tau}  }} (p_1+p_{2};  p_3, p_4,
\cdots, p_{n+1}) \big |^2
\\ &
+ \delta \int d \mu_{n+1} (\pb_{n+1}) \;
\Thnone { |e^{i r_3}- e^{ - i r_{4}}|^2 }
\\ &
\qquad  \times \big |{F}_\good^{ 2 \tau}  (p_3+p_{4}; p_1, p_2, p_5,
\cdots, p_{n+1}) \big |^2
\end{align*}
Both terms can be estimated   by similar arguments used for $Q_1$.
So we obtain \eqref{off2}.
\myendproof

\bigskip

\section{Conclusions}
From the main estimate Theorem \ref{mainest}, we need the
relation
$$
\kappa_{n-1} =  1- \kappa_n/2.
$$
To satisfy this relation,
for any large integers $N$ fixed, we let
\begin{equation}\label{c0}
\kappa_{n}= 2/3+ (-1)^n 2^{-2N+n}/3  , \quad n =1, \cdots, 2N+1.
\end{equation}
A few terms are given explicitly in the following:
$$
\kappa_{2 N+1}= 2/3-2/3= 0, \quad \kappa_{2N}=2/3+1/3=1, \quad
\kappa_{2 N-1}= 2/3-1/6,
$$
$$
\kappa_{2 N-2}= 2/3+1/12, \quad \cdots, \; \kappa_{2}= 2/3 +
 2^{-2N+2}/3
$$

We first apply Theorem \ref{mainest}  to have
$$
A_+^\ast D_{2N+1}^{-1} A_+ \le
C \loggl^2 \U^{2N}_{\kappa_{2N},  \tau}
$$
In order to satisfy the condition $\gamma \le \loggl^{-3}$ later on,
we now replace $\loggl^2$ on the right side by $\loggl^3$ to have a further
upper bound.
Now we apply the lower bound part of  Theorem \ref{mainest} to have
$$
A_+^\ast \bigg \{ D_{2N}+ A_+^\ast D_{2N+1}^{-1} A_+ \bigg \}^{-1} A_+ \ge
C \loggl^{-3} \V^{2N-1}_{\kappa_{2N-1},  2\tau}
$$
We can repeat this procedure until we have
$$
A_+^\ast \big ( D_3 + \cdots \big ) ^{-1}  A_+
\le C  \loggl^{2N+4} \U^{2}_{\kappa_{2}, \tau}
$$
Thus we have
$$
\l w, \Big [ D_{2}  + A_+^\ast \big ( D_3 + \cdots \big ) ^{-1}  A_+ \Big
]^{-1} w \r \ge
\l w, [ D_2+ C  \loggl^{2N+4} \U^{2}_{\kappa_{2},  \tau}]^{-1}  w \r
$$

The Fourier transform of $w$ is
$$
\hat w (p_1, p_2)= e^{-i r_2}
$$
Since $p_1+p_2=0$ under the measure $d \mu_2$, we have
\begin{align*}
& \l w, [ D_2+ C  \loggl^{2N+4} \U^{2}_{\kappa_{2},  \tau}]^{-1}  w \r \\
=&  \frac 1 2 \int\;  dp_1 \;  \Big \{ \lambda+ 2\omega (p_1)+
C  \loggl^{2N+4} \U^{2}_{\kappa_{2},  \tau}(p_1, -p_1)  ]\big | \Big \}^{-1}
\end{align*}

The last integration is the same as the right side of \eqref{low0} with
$n=1$ and
$A_+ F$ replaced by one. Following similar argument,
we have
$$
\l w, [ D_2+ C  \loggl^{2N+4} \U^{2}_{\kappa_{2},  \tau}]^{-1}  w \r \\
\ge
C \loggl^{-2N-4} K^\tau_{\kappa_2}(0,0)
$$
where $K^\tau_{\kappa_2}(0,0)$ is defined in Lemma \ref{le:up}.
From \eqref{eq:lowest}, we have
$$
K^\tau_{\kappa_2}(0,0) \ge \logl^{\kappa_1}\, , \qquad \kappa_{1}= 2/3 -
 2^{-2N+1}/3
$$
Thus we have
$$
\l w, [ D_2+ C  \loggl^{2N+4} \U^{2}_{\kappa_{2},  \tau}]^{-1}  w \r \\
\ge
\loggl^{-2N-4} \logl^{\kappa_1}
$$
Therefore, we have the lower bound
\begin{align*}
& \l w, \Big [ D_{2}  + A_+^\ast \big ( D_3 + \cdots \big ) ^{-1}  A_+ \Big
]^{-1} w \r  \\
& \ge \logl^{2/3}
\exp \Big [ \,  - \frac  \loggl { 2^{2N-1} 3 } - (2N+4)\logggl \, \Big ]
\end{align*}
By  choosing
$$
N = \alpha \logggl
$$
with  $\alpha$ large enough, together with Lemma \ref{le:mono}
we have proved the lower bound.

\bigskip
Instead of \eqref{c0}, we can choose
$$
\kappa_{n}= 2/3- (-1)^n 2^{-2N+n+1}/3  , \quad n =1, \cdots, 2N.
$$
Explicit examples are
$$
\kappa_{2 N}= 2/3-2/3= 0, \quad \kappa_{2N-1}=2/3+1/3=1, \quad
\kappa_{2 N-2}= 2/3-1/6,
$$
$$
\kappa_{2 N-3}= 2/3+1/12, \quad \cdots, \; \kappa_{2}= 2/3 -
2^{-2N+3}/3
$$
With this choice of $\kappa_n$, similar argument proves the upper
bound.
This concludes Theorem \ref{resest}.
\bigskip

\noindent{\bf Acknowledgement}:
I would like to thank  P. Deift, J. Baik
and H. Spohn for explaining their results to me.
In particular, Spohn has pointed out the relation
\eqref{0.1} so that the connection between the current
across the zero and the diffusion coefficient becomes
transparent. I would also like to thank A. Sznitman for
his hospitality and invitation to lecture on this
subject at ETH.

\bigskip
\bigskip

\noindent{Horng-Tzer Yau}, yau@cims.nyu.edu \\
 Courant Institute, New York University, 251 Mercer Street, New York, NY 10012, USA

\end{document}